\documentclass[10pt,apl,aip,twocolumn,showpacs,preprintnumbers,boldmath,amsmath,amssymb,floatfix]{revtex4-1}

\usepackage{times}
\usepackage{graphicx}
\usepackage{amssymb}
\usepackage{amsmath}
\usepackage{amsfonts}
\usepackage{dcolumn}
\usepackage{braket}
\usepackage{bm}
\usepackage{multirow}
\usepackage[colorlinks=true,
            linkcolor=red,
            urlcolor=blue,
            citecolor=blue]{hyperref}
\usepackage{amstext}
\usepackage{framed}
\graphicspath{{Images/}}
\DeclareGraphicsExtensions{.pdf,.eps,.png}

\begin{document}
\title{Defect induced polarization and dielectic relaxation in Ga$_2$$_-$$_x$Fe$_x$O$_3$}

\author{Sudipta Mahana}
\affiliation{Institute of physics, Sachivalaya Marg, Bhubaneswar - 751005, India}
\affiliation{Homi Bhabha National Institute, Training School Complex, Anushakti Nagar, Mumbai - 400085, India}

\author{C. Dhanasekhar}
\affiliation{Department of physics, Indian Institute of Technology, Kharagpur -721302, India }
\affiliation{Cryogenic Engineering Centre, Indian Institute of Technology, Kharagpur - 721302, India}
\author{A. Venimadhav}
\affiliation{Cryogenic Engineering Centre, Indian Institute of Technology, Kharagpur - 721302, India}
\author{D. Topwal}
\email{dinesh.topwal@iopb.res.in, dinesh.topwal@gmail.com}
\affiliation{Institute of physics, Sachivalaya Marg, Bhubaneswar - 751005, India}
\affiliation{Homi Bhabha National Institute, Training School Complex, Anushakti Nagar, Mumbai - 400085, India}

\date{\today}

\begin{abstract}

We report results of the dielectric and pyroelectric measurements on solid solutions of Ga$_2$$_-$$_\emph{x}$Fe$_\emph{x}$O$_3$ with $\emph{x}$ = 0.75, 1.0 and 1.25.  These systems exhibit dipolar cluster glass behavior in addition to the spin glass behavior making them belong to a class of few systems showing multiglass behavior. Presence of two contributing relaxations in dielectric data are observed possibly due to the flipping and breathing of polar nano-clusters. Further, emergence of polarization in these systems can be understood in terms of thermally stimulated depolarization current (TSDC) effect caused by defect dipoles possibly associated with charged oxygen vacancies rather than the intrinsic ferroelectric behavior.

\end{abstract}

\maketitle


 Mutiferroics have gained considerable attention because of their scientific interest and potential for technological applications in magnetic sensors, multi-state memories, etc \cite{eerenstein2006multiferroic}. Mostly, multiferroics have ferroelectric transitions either at very low or very high temperatures. However, for practical applications, it is of prime importance to establish such properties in the vicinity of room temperature. \\
GaFeO$_3$ is expected to be ferroelectric because of the non-centrosymmetric ${Pc}$2${_1}$$n$ crystal symmetry \cite{mahana2017complex}. It has three octahedral cation sites (Fe1, Fe2, Ga2) and one tetrahedral site (Ga1) and possess inherent site disorder caused by similar ionic radii of Ga and Fe which results in its ferrimagnetic type behavior \cite{mahana2017complex} and is believed to have very large polarization \cite{stoeffler2012first,stoeffler2013first}. Several groups have studied polarization vs electric field ($P$-$E$) response in GaFeO$_3$, mostly using thin-films and found very small coercieve field ($E_c$) and remnant polarization ($P_r$), inconsistent with the theoretical calculations \cite{trassin2007epitaxial,sharma2013study,oh2015room}. Mukherji ${et}$ ${al.}$ argued the presence of room temperature nanoscale ferroelectricity from their observation of a 180$^0$ phase shift in the piezoresponse \cite{mukherjee2013room}. However, this phase shift does not necessarily indicate ferroelectric (FE) behavior. On the contrary, Song ${et}$ ${al.} $\cite{song2016ferroelectric} achieved very large $E_c$ and $P_r$ by applying very high bias electric field, which is in good agreement with ab-initio calculations \cite{stoeffler2012first,stoeffler2013first} . However, presence of ferroelectricity in bulk GaFeO$_3$ is rather inconclusive due to leaky behavior of the sample \cite{naik2009electrical}. Saha $et$ $al.$ \cite{saha2012multiferroic} studied FE property in bulk GaFeO$_3$ by pyroelectric measurements and showed the emergence of polarization below magnetic transition. They claimed from symmetry analysis that non-centrosymmetric magnetic ordering coupled with inherent site-disorder below $T_C$  drives ferroelectricity in GaFeO$_3$. Note, pyroelectric measurements showing spontaneous polarization does not necessarily indicate intrinsic FE behavior, pyrocurrent (polarization) can also arise from dipole reorientation and release of charges from localized states as observed in Y$_3$Fe$_5$O$_{12}$ \cite{kohara2010excess}, YFe$_{0.8}$Mn$_{0.2}$O$_3$ \cite{cho2017absence} and various manganites \cite{zhang2014investigation,zou2014excess} etc. Therefore, it is essential to explore the differences of pyroelectric currents related to FE phase transition and that induced by dipole reorientation or the release of charges from localized states i.e. thermally stimulated depolarization current (TSDC) effect.\\
In this study, we discuss in detail pyroelectric measurements in solid solutions of  Ga$_2$$_-$$_\emph{x}$Fe$_\emph{x}$O$_3$ ($\emph{x}$ = 0.75, 1.0 and 1.25) to elucidate the origin of ferroelectric polarization. Our results indicate that induced polarization is associated with TSDC effect rather than the true FE behavior. We have also discussed dielectric relaxations behavior of these systems and their correlations with TSDC effect.


The polycrystalline samples of Ga$_2$$_-$$_\emph{x}$Fe$_\emph{x}$O$_3$ ($\emph{x}$ = 0.75, 1.0 and 1.25) were prepared by solid-state route and details are described elsewhere \cite{mahana2016giant,mahana2017complex}. Detailed structural and magnetic characterizations has been reported earlier \cite{mahana2017complex}. Temperature dependent dielectric measurements were performed using an impedance analyser 4291A in the frequency range of 1 kHz -1 MHz, under dc bias voltage of 1 V. The pyroelectric current ($I$) was measured using a Keithley 6517A electrometer and was recorded during the heating process after sample was cool down under a poling field ($E_p$ = $\pm$ 1.1 kV/cm) from high temperature and was short-circuited for long time to remove any residual charges. Spontaneous electric polarization was obtained by integrating the current with respect to time. \\

Figure 1 (a-c) depict temperature dependence of the dielectric constant ($\varepsilon$$^\prime$($T$)) of  Ga$_2$$_-$$_\emph{x}$Fe$_\emph{x}$O$_3$ ($\emph{x}$ = 0.75, 1.0 and 1.25) at different frequencies ranging from 1 kHz to 1 MHz. $\varepsilon$$^\prime$($T$) shows frequency-dependent dielectric relaxations for all the compounds . The dielectric relaxation peaks are found to gradually shift to higher temperature with increasing frequency. Two dielectric relaxations, A and B, are clearly visible, for $x$ = 1.0 and 1.25 compositions, where as a single peak is observed for $x$ = 0.75 compound in the limited measurement window. Such an observation suggests a combined relaxation mechanism like Maxwell-Wagner relaxations (which originate from the presence of accumulated charge carriers between regions in the sample that have different conductivities such as near the grain boundaries) and Debye relaxations (arise from dipolar contributions originating from the asymmetric hopping of charge carriers in the presence of an electric field) in the proposed system, as reported earlier in La$_2$NiMnO$_6$ \cite{choudhury2012near}. The presence of strong dielectric relaxations is also observed in
the corresponding dielectric loss ($D$ = tan$\delta$) data as shown in Fig. 1 (d-f). To have a better understanding of the origin of observed dielectric relaxations behavior in these compounds, we have systematically analyzed the dielectric data. \\
Frequency dependence of dielectric loss ($D$) peak diverges according to critical power law given by following equation, \cite{souletie1985critical}
\begin{eqnarray}
   \tau = \tau_0 (\frac{T_m - T_f}{T_f})^{- z \nu},  
\end{eqnarray}

where $T_m$ is the peak temperature of dielectric loss curve, $T_f$ is the freezing temperature, $z$ is the dynamic critical exponent, $\nu$ is the critical exponent of the correlation length and $\tau_0$ (1/$f_0$) is the shortest relaxation time available to the system, i.e. the microscopic flipping time of fluctuating entities. This is a typical characteristic of the electric glass behavior. Fitting of frequency dependence of  $T_m$  with critical power law for Ga$_2$$_-$$_\emph{x}$Fe$_\emph{x}$O$_3$ ($\emph{x}$ =  1.0 and 1.25) are shown in Fig. 2 (a) and (b), respectively and the best fitted parameters are summarized in Table I.  The parameters obtained signifies dipolar cluster glass behavior, as observed in BaTi$_{0.7}$Sn$_{0.3}$O$_3$ \cite{kleemann2014non} and Fe$_2$TiO$_5$ \cite{sharma2014multiglass} and are associated with the formation of polar nano-regions (PNRs) and interactions among them. Earlier we have reported, the existence of spin glass behavior in these systems \cite{mahana2017complex}. Thus Ga$_{2-x}$Fe$_x$O$_3$ are unique systems exhibiting multiglass behavior (both spin glass and dipolar glass), as observed  in very few systems like Fe$_2$TiO$_5$ \cite{sharma2014multiglass}, SrTi$_{0.98}$Mn$_{0.02}$O$_3$ \cite{shvartsman2008sr} and CuCr$_{1-x}$V$_x$O$_2$ \cite{kumar2013spin}. \\

We analyzed dielectric relaxation behavior using Vogel-Fulcher (V-F) law, which probes the dynamics and population profile of the dipolar responses as a function temperature. This includes an increasing degree of interaction between random local relaxation processes, given by following expression,
\begin{eqnarray}
f =f_0 exp[{\frac{-E_a}{k_B(T_m-T_v)}}]
\end{eqnarray} 
\begin{figure}[!ht]
 \centering
 \includegraphics[height=6.5cm,width=8.5cm]{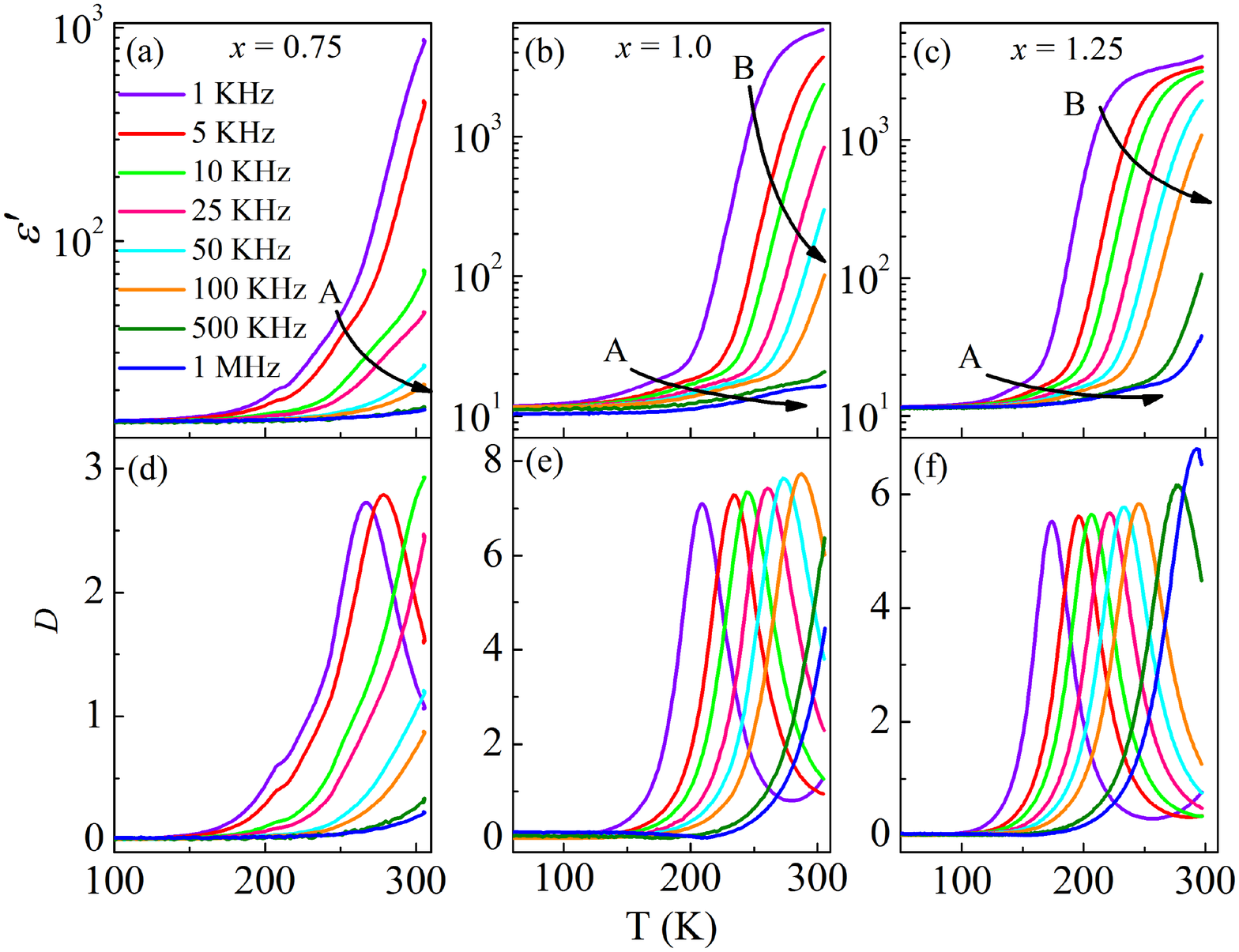}
 \caption{(a), (b) and (c) Temperature dependent dielectric constant ($\epsilon$$^{prime}$($T$)) measured at several frequencies of Ga$_2$$_-$$_\emph{x}$Fe$_\emph{x}$O$_3$, for $\emph{x}$ = 0.75 (a), 1.0 (b) and 1.25 (c). Corresponding dielectric loss ($D$ = tan$\delta$) are shown in (c), (d) and (f), respectively.}
 \label{fig}
 \end{figure}
Here $E_a$ is the activation energy or the potential barrier separating different metastable states accessible to the system, $f_0$ is the characteristic frequency, $T_m$ is the peak temperature of dielectric loss curve and $T_v$ is the Vogel-Fulcher temperature, which measures interparticle or intercluster interaction strength. Interestingly, it is observed that whole region is not fitted well with V-F law for both the compounds. High temperature region (high frequency) is fitted well with the V-F law, where as the low temperature (low frequency) region is fitted well with the Arrhenius law, as given by,\\
\begin{eqnarray}
 f = f_0 exp[\frac{-U}{k_BT_m}], 
\end{eqnarray} 
where $f_0$ is the characteristic frequency and $U$ is the activation energy. The fitting of data with the V-F law (solid line) and Arrhenius law (dashed line) for $x$ = 1.0 and 1.25 compounds are shown in Fig. 2 (c) and 2 (d), respectively. 
The best fitting parameters obtained from the fitting with the V-F and Arrhenius laws are listed in Table I and are comparable with the parameters listed in literature. \\
\begin{figure}[!ht]
 \centering
 \includegraphics[height=6.5cm,width=8.5cm]{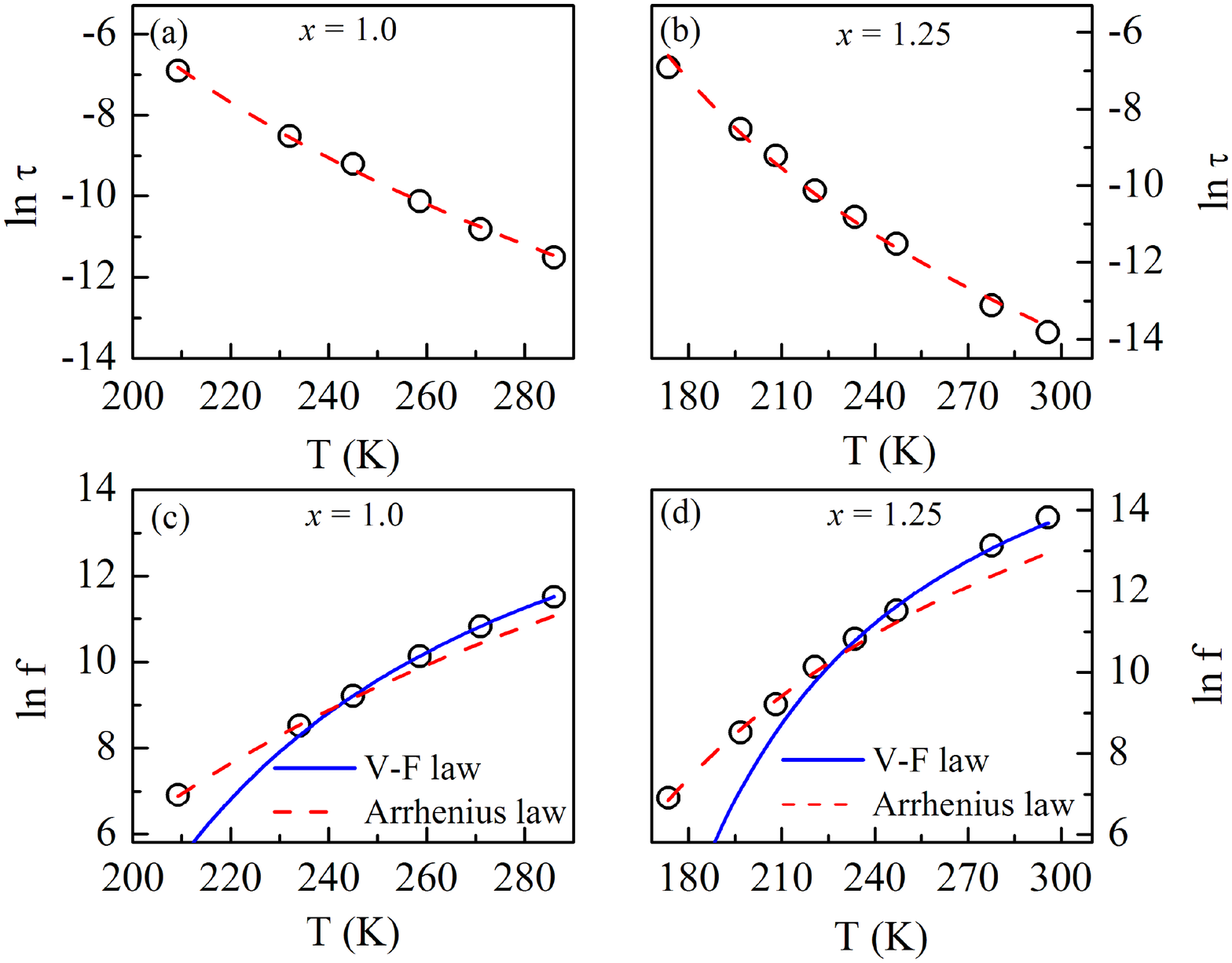}
 \caption{Plot of ln$\tau$ vs. peak temperature ($T$) obtained from the peak of  loss ($D$), with dashed line representing the best fit with critical power law of Ga$_2$$_-$$_\emph{x}$Fe$_\emph{x}$O$_3$, for $x$ = 1.0 (a) and 1.25 (b). Plot of ln$\tau$ vs. $T$ with solid line representing the best fit with Vogel-Fulcher law and dashed line representing the best fit with Arrhenius law of Ga$_2$$_-$$_\emph{x}$Fe$_\emph{x}$O$_3$, for $x$ = 1.0 (c) and 1.25 (d). }
 \label{fig:xrd.pdf}
 \end{figure}
Such two overlapping relaxation contributions were also observed in Sr$_2$LaTi$_2$Nb$_3$O$_{15}$ \cite{bovtun2007relaxor} and various lead-based relaxor ferroelectrics with perovskite structure \cite{macutkevic2006infrared,bovtun2009broadband}, where in two contributing relaxations in the broad dielectric loss peak were attributed to a breathing and flipping mechanism of PNRs. On the contrary, Rotaru ${et}$ ${al.}$ \cite{rotaru2011origin} reported that in Ba$_6$ScNb$_9$O$_{30}$, dielectric constant data were best described by the V-F expression while the dielectric loss displayed Arrhenius behavior, which were attributed to the presence of  two separate relaxation processes: one dominates the dielectric polarizability and the other dominates the loss . In Ga$_2$$_-$$_\emph{x}$Fe$_\emph{x}$O$_3$ the first scenario can be rationalized as showing similar behavior as that observed in Sr$_2$LaTi$_2$Nb$_3$O$_{15}$ \cite{bovtun2007relaxor} (two temperature regions with two relaxation behaviors), implying the presence of two relaxation contributions to the dielectric  response, which are not resolvable. At high temperature region the flipping of the polar nano-clusters i.e. dipole reversal is the dominant mechanism. On cooling the interaction among the clusters increases and causes freezing of local dipole moments (at least partially), when approaching the freezing temperature $T_f$. Thereafter, flipping mechanism gets suppressed and the breathing mechanism (fluctuations in the cluster walls, or, in other words, cluster volume fluctuation) become dominant below $T_f$ . Flipping concerns the central part of each cluster only, but their walls (probably relatively broad) are apparently still actively vibrating. Moreover, a broad frequency and temperature range dielectric study is required for clearly understanding the relaxation behavior. \\

It is observed that the obtained value of  $\tau_0$ using V-F law is almost three orders of magnitude smaller than that obtained using critical power law. Such a difference (also seen in various glassy systems) was explained by Souletie and Tholence \cite {souletie1985critical}, who showed that high temperature expansion of power law is identical to V-F law up to terms of order of ($T_0$/$T$)$^3$  and therefore, provides a discrimination in the fitting parameters using these laws.The obtained activation energies for dielectric relaxations (both from Arrhenius behavior and V-F law) are different from that obtained for the magnetic relaxation process \cite{mahana2017complex}, suggesting that both processes have different origins. Inherent site-disorder coupled with competing magnetic interaction is the origin of spin glass behavior in these compounds \cite{mahana2017complex} and the activation energies obtained from dielectric relaxations signify possibility of defect induced polaron hopping process which will be discussed later. \\

\begin{figure}[!ht]
 \centering
 \includegraphics[height=6.5cm,width=8.5cm]{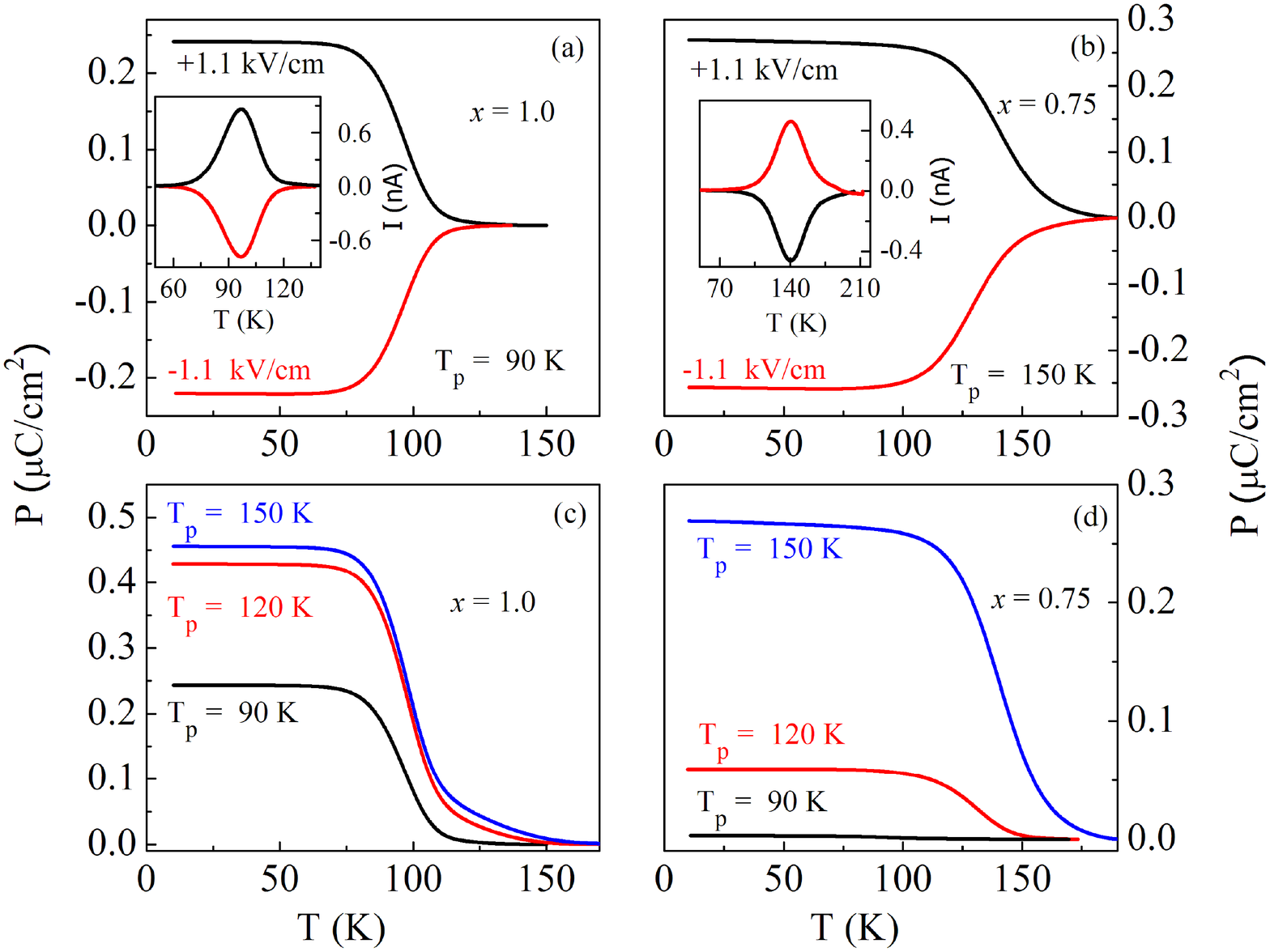}
 \caption{Temperature dependent polarization of Ga$_2$$_-$$_\emph{x}$Fe$_\emph{x}$O$_3$, for $x$ = 1.0 (a) and 0.75 (b). Corresponding pyroelectric current peaks are shown in the insets. Temperature dependent polarization at different poling temperatures of Ga$_2$$_-$$_\emph{x}$Fe$_\emph{x}$O$_3$, for $x$ = 1.0 (c) and 0.75 (d). }
 \label{fig:xrd.pdf}
 \end{figure}
 Further we investigated polarization properties in Ga$_2$$_-$$_\emph{x}$Fe$_\emph{x}$O$_3$. Pyroelectric current measurements were performed at various poling temperatures ($T_p$) and also with different heating rates keeping fixed poling field ($E_p$) and $T_p$. Due to the high leakage current pyroelectric measurement on $x$ =1.25 composition could not be performed. Fig. 3 (a) depicts temperature dependence of pyroelectric current and polarization measurements for GaFeO$_3$ under a $E_p$ of 1.1 kV/cm and $T_p$ of 90 K, in which polarization starts to emerge below 120 K and it is reversible under reversed $E_p$. Our results are in agreement with earlier study reported by Saha ${et}$ ${al.}$ \cite{saha2012multiferroic}, where in they attributed that non-centrosymmetric magnetic ordering coupled with inherent site-disorder drives ferroelectricity in this system. We also present results for $x$ = 0.75 compound Fig. 3 (b) for a $T_p$ of 150 K. Such polarization curves are often interpreted as imprints of FE behavior \cite{sharma2014multiglass}. It has to be noted that for $x$ = 0.75 composition, polarization emerges well above $T_c$, suggesting that it can not be driven by magnetic ordering as proposed by Saha ${et}$ ${al.}$ for GaFeO$_3$ \cite{saha2012multiferroic}. Moreover, the emergence of pyroelectric current can not be ascribed to the intrinsic FE behavior because no anomaly is observed in $\epsilon$$^\prime$($T$) (Fig. 1) and heat capacity data (not shown here) at the pyroelectric peak temperature \cite{kohara2010excess, cho2017absence,zhang2014investigation,zou2014excess}. The observed phenomena could however be understood in terms of TSDC, which mainly originates from the thermal release of trapped-charges and gives rise to current peak upon heating \cite{raju2016dielectrics}. The correlation between TSDC and polarization is clearly seen in Fig 4(c) and 4 (d), which show an increase of  transition temperature and polarization with increase of poling temperature.\\

Further analysis was performed to understand the TSDC effect. In the Bucci-Fieschi-Guidi framework, which assumes that a system has a single relaxation time ($\tau$($T$)) and the decay rate of Polarization ($P$) that means pyroelectric current ($I$) is described by -$P$/$\tau$($T$).  This leads to a correlation between pyroelectric current, Polarization, relaxation time and activation energy, which is described by the following relation, \cite{kohara2010excess}

\begin{eqnarray}
ln\frac{P(T)}{I(T)} = \frac{E}{k_BT} + ln\tau_0,
\end{eqnarray}  
where $T$ is the temperature, $E$ is the activation energy and $\tau_0$ is the shortest relaxation time. Plot of log($P$/$I$) vs the inverse temperature (1/$T$) along with its fitting with Eq. (4) for $x$ = 1.0 and 0.75 compositions are shown in Fig. 4 (a) and (b), respectively. The best fitting parameter obtained are $E$ = 0.08 eV and $\tau_0$ = 7.01 $\times$ 10$^{-5}$ s for $x$ = 1.0 compound and those for $x$ = 0.75 composition these parameters are 0.11 eV and 6.11 $\times$ 10$^{-6}$ s, respectively.\\

The dependency of pyroelectric peak temperature ($T_m$) on heating rate ($b$ = ${dT}$/${dt}$) for TSDC process is governed by the following equation,\cite{kohara2010excess}

\begin{eqnarray}
ln\frac{T_m^2}{b} = \frac{E}{k_BT_m} + ln\frac{\tau_0E}{k_B}
\end{eqnarray}  
                       \begin{figure}[!ht]
 \centering
 \includegraphics[height=6.5cm,width=8.5cm]{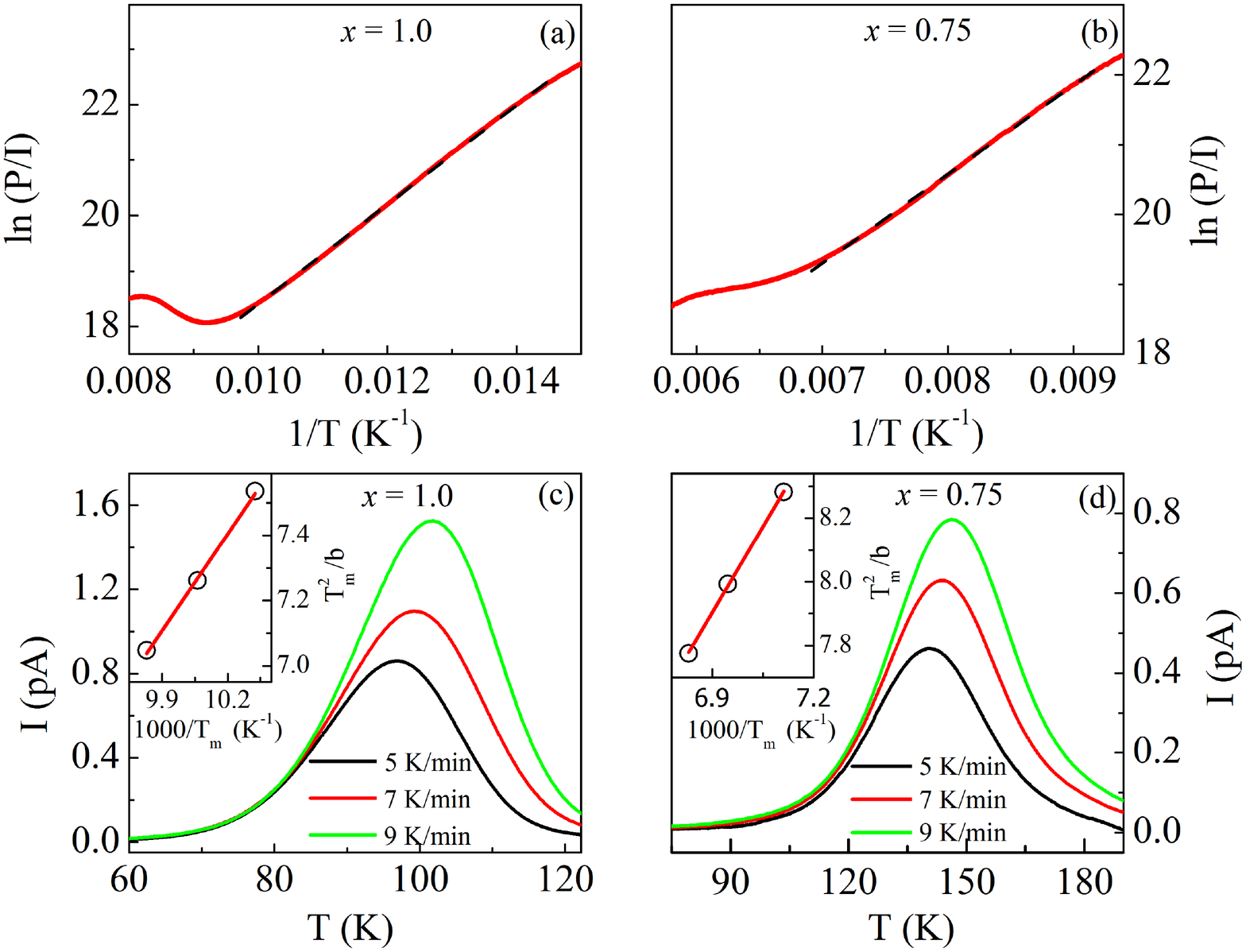}
 \caption{Plot of ln($P$/$I$) vs. inverse of temperature, with dashed line representing the best fit with Eq. (4) of Ga$_2$$_-$$_\emph{x}$Fe$_\emph{x}$O$_3$, for $x$ = 1.0 (a) and 0.75 (b). Variation of pyroelectric peak with heating rate of Ga$_2$$_-$$_\emph{x}$Fe$_\emph{x}$O$_3$, for $x$ = 1.0 (c) and 1.25 (d). Corresponding insets represent fitting with Eq. (5), which describes relation between pyroelectric peak temperature ($T_m$) and heating rate ($b$).}
 \label{fig:xrd.pdf}
 \end{figure}  
It is expected that $T_m$ shifts to high temperature with increasing heating rates as per Eq. (5) and it is also observed for the present systems as shown in Fig. 4 (c) and (d). This also confirms TSDC nature of the pyroelectric current instead of intrinsic FE behavior where no such temperature dependence is expected \cite{zhang2014investigation}. Insets of  Fig. 4 (c) and (d) show the fitting of  heating rate dependence of $T_m$ with Eq. (5). The best fitting gives $E$ = 0.08 eV and 0.15 eV for $x$ = 1.0 and 1.25 compounds, respectively. The activation energy obtained from TSDC effect agrees well with that obtained from V-F law fitting of dielectric data implying that dipolar cluster glass behavior and polarization emerged are correlated. It also suggests origin of the polarization is from the breathing of the frozen polar regions in the materials. As FE is not an intrinsic behavior, defect-induced dipoles create PNRs and induce electric glassy behavior and polarization in these systems.  The obtained activation energy (0.09 eV) is same as that obtained from bulk relaxation in single crystal of GaFeO$_3$ reported by Mukherjee ${et} {al.}$ \cite{mukherjee2013dielectric}. They attributed this to an interplay between lattice and polaronic defects associated with charged oxygen vacancies, which are generally formed during high temperature synthesis process as seen in various systems \cite{mukherjee2013dielectric,ke2010oxygen} as per Kroger-Vink notation, \cite{kroger1956relations} given by the following reactions:\\

     \qquad                O$_o$ $\rightarrow$ V$_o^\times$ $+$ $\frac{1}{2}$O$_2$$\uparrow$ \\

       \qquad             V$_o^\times$ $\rightarrow$ V$_o^\bullet$ $+$ e$^\prime$\\

          \qquad          V$_o^\bullet$ $\rightarrow$ V$_o^{\bullet\bullet}$ $+$ e$^\prime$ \\

The electrons associated with this charged oxygen vacancy are localized and could only hop to neighboring sites and results in reorientation of defect dipoles. 
These defect dipoles start interacting via PNRs formation and freeze out when the temperature approaches the freezing point $T_f$ resulting in the observed electric polarization. \\

 In summary, the detailed dielectric studies show two relaxation processes. The higher frequency component is assigned to flipping of PNRs while the lower-frequency component to their breathing. The flipping relaxation frequency follows the Vogel-Fulcher law with freezing near $T_f$ and thereafter breathing mechanism dominates satisfying the Arrhenius law. Further, pyroelectric study confirmed that emergence of polarization in bulk  Ga$_2$$_-$$_\emph{x}$Fe$_\emph{x}$O$_3$ can be understood in terms of TSDC effect rather than the intrinsic ferroelectric behavior. It is seen that TSDC effect is correlated with the dipolar cluster glass behavior in the system caused by defect induced dipoles possibly associated with charged oxygen vacancies.
\begin{table}[]
\centering
\caption{Various physical parameters obtained from the fitting of frequency dependence of peak temperature with Critical-power law, V-F law and Arrhenius law  for Ga$_2$$_-$$_\emph{x}$Fe$_\emph{x}$O$_3$ ( $x$ = 1.0 and 1.25).}
\begin{tabular}{m{2.6cm}  m{1.5cm}  m{2cm} m{2cm} m{2cm}  }
    \hline
          Composition ($x$) & &1.0 & 1.25  \\[1ex]
           \hline

 & $z\nu$& 6.23 & 6.88 \\[1ex]
    Critical-power law       & $\tau_0$ (s) & 7.14$\times$ 10$^{-5}$ & 1.23$\times$ 10$^{-6}$ \\[1ex]
         & $T_{g}$ (K) & 125 & 108   \\[1ex]
 \hline
         & $E_a$ (eV) & 0.091 & 0.093 \\[1ex]
      V-F law     & $\tau_0$ (s)  & 1.34$\times$ 10$^{-8}$ & 3.4$\times$ 10$^{-9}$ \\[1ex]
         & $T_{VF}$ (K) & 127 & 107   \\[1ex]
\hline
         Arrhenius law & $U$ (eV) & 0.28 & 0.22 \\[1ex]
          & $\tau_0$ (s) & 1.81$\times$ 10$^{-10}$ & 4.32$\times$ 10$^{-10}$   \\[1ex]

      \hline
   \end{tabular}
   \label{tab:example}
\end{table}

\bibliographystyle{apsrev4-1}
\bibliography{ref}

\end{document}